\documentclass[]{interact}

\usepackage{epstopdf}
\usepackage[caption=false]{subfig}

\usepackage[numbers,sort&compress,merge]{natbib}
\bibpunct[, ]{[}{]}{,}{n}{,}{,}
\renewcommand\bibfont{\fontsize{10}{12}\selectfont}

\theoremstyle{plain}

\theoremstyle{definition}

\theoremstyle{remark}

\begin{document}


\title{EMReact: A Tool for Modelling Electromagnetic Field Induced Effects in Chemical Reactions by Solving the Discrete Stochastic Master Equation}

\author{
\name{Kelvin Dsouza\textsuperscript{a} and Daryoosh Vashaee\textsuperscript{a,b}\ddag\thanks{\ddag Corresponding author: dvashae@ncsu.edu}}
\affil{\textsuperscript{a}Electrical and Computer Engineering Department, North Carolina State University, Raleigh, North Carolina, USA}
\affil{\textsuperscript{b}Materials Science and Engineering Department, North Carolina State University, Raleigh, North Carolina, USA}
}

\maketitle

\begin{abstract}
The effects of electromagnetic fields (EMF) have been widely debated concerning their role in chemical reactions. Reactions, usually took hours or days to complete, have been shown to happen a thousand times faster using EMF radiations. This work develops a formalism and a computer program to evaluate and quantify the EMF effects in chemical reactions. The master equation employed in this program solves the internal energy of the reaction under EMFs while including collisional effects. Multiphoton absorption and emission are made possible with the transitioning energy close to the EMF and are influenced by the dielectric properties of the system. Dimethyl Sulfoxide and Benzyl Chloride are simulated under different EMF intensities. The results show that EMF absorption is closely related to the collisional redistribution of energy in molecules. The EMF effect can be interpreted as a shift of the thermodynamic equilibrium. Under such nonequilibrium energy distribution, the "temperature" is not a reliable quantity for defining the state of the system.
\end{abstract}


\renewcommand{\abstractname}{Program Summary}

\begin{abstract}
\textit{Program Title:} EMReact\\
\textit{Developer's repository link:} https://github.com/keludso/EMReact.git\\
\textit{Licensing provisions:} BSD License\\
\textit{Programming language:} MATLAB\\
\textit{Nature of problem:} The role of Electromagnetic radiation in chemical reactions.\\
\textit{Solution method:} Solving the Discrete Energy Stochastic Master Equation employing the Gillespie’s Algorithm. 
\end{abstract}

\section{Introduction}

Electromagnetic field-assisted processes in the form of microwave heating span a wide range of research and industrial areas, such as low-temperature sintering, low-temperature de-crystallization, enhancement of reaction rates, and catalytic effects in organic and inorganic synthesis\cite{Amin2018_1,vashae2016_2,Amin2019_3,Amin2020_4,RYBAKOV20199567_5, Nuchter2003_6,Horikoshi2018_7,Kappe2004-nw_8,MW2017_9,Danks2018_10}. While the source of the heating mechanism is well understood, certain effects such as enhancement of reaction rates, increased yield, and creation of distinct transition pathways, which cannot be achieved through the conventional heating mechanism, have led to the term non-thermal or EMF specific effects\cite{A827213Z_11,Herrero2007-vx_12,Shibata-1996_13,Garbacia2003_14,Kanno2012-qo_15,Jacob1995_16,Rustum2002_17}. The term non-thermal effect is loosely described as any reaction or process that cannot be achieved through the conventional heating technique, which has gained a reputation for being something mysterious. There has been an increasing debate among the scientific community regarding the non-thermal effects being just a manifestation of the thermal effect in the form of non-uniform localized heating, hotspots, and/or inaccuracy in temperature measurement\cite{Kappe2013_18}.  Exhaustive reviews have been published on both the thermal and non-thermal effects\cite{B411438H_19,Spargo2005_20,Bogdal2017_21}. While there are satisfactory explanations for the observed phenomena in each case, modeling these effects to get an intuitive understanding of the process remains challenging. 

The microwave energy is typically too low to cause electronic transitions in materials, and unlike the laser, microwave absorption has a more subtle effect on the system. The microwave field interacts with the dipole of the molecule, increasing its energy. The thermal effect in microwave absorption can be explained due to energy lost to the system due to the relaxation of these excited dipole moments. By tracking the internal vibrational energy distribution of the system, we can get information regarding the equilibrium properties and temperature of the system. We can model microwave absorption as multiphoton absorption in a collisional environment to implement this idea. Intramolecular vibrational redistribution and vibrational energy transfer can be used to model energy dissipation among vibrational levels\cite{Stannard1981_22,A704144F_23}. As we will see, collisional effects play a vital role in the redistribution of energy, which gives rise to the heating of the sample as an observable quantity. 

Ma J., in their paper, used internal energy distribution to explain non-thermal microwave effects using a two-channel process where he hypothesized that microwave activates a rapid channel to enhance the reaction rate demonstrating the distribution of internal states deviate from equilibrium under a strong microwave field\cite{Ma2016_24, Jianyi_2021}. The non-thermal microwave effect in certain applications has been explained by the (1) increase in pre-exponential factor in the Arrhenius Equation, which directly influences the molecular collisions, (2) Decrease in the activation energy, which directly increases the reaction rate, (3) Localized microscopic high temperatures (4) Intermediate transition states, and (5) polar specific effects in the medium\cite{PERREUX20019199_25} . 

This paper introduces a model using the master equation formulation to demonstrate the effect of the microwave as a multiphoton absorption in a collisional environment.  

\section{Methodology}

\subsection{Master Equation}
The temporal evolution of the population of molecules under thermal dissociation, including the multiphoton absorption, can be described by the energy-grained master equation:\cite{SHIN1994105_26,Toselli_27} 

\begin{equation}
\begin{aligned}
 \frac{dN_i}{dt}=\frac{I(t)}{\hbar\omega}[\sigma_{i,i-1}N_{i-1}+\frac{\rho_i}{\rho_{i+1}}\sigma_{i+1,i}N_{i+1}-(\sigma_{i+1,i}+\frac{\rho_i}{\rho_{i+1}}\sigma_{i,i-1})N_i]\\
 +\omega\sum_{j}P_{i,i+1}N_i-\omega\sum_{j}P_{i,i-1}N_i-\sum_{m}k_mN_i
\end{aligned}
\end{equation}

where the concentration of the species at the i\textsuperscript{th} energy level is given by $N_i$. I(t) is the intensity (W/cm\textsuperscript{2}) of the microwave radiation. $P_i$  is the transfer probability of the molecule from the i\textsuperscript{th} energy level. $\omega$ is the collisional frequency. $k_i$  is the unimolecular rate constant of the ith channel. $\sigma_{i+1,i}$ is the microscopic cross-section for the absorption of microwave energy from level i to i+1. $\rho_i$ is the density of states at the i\textsuperscript{th} energy level. 

As it is difficult to find a solution for the Master equation given in Eq(1), a stochastic algorithm is used to solve it, as provided by Gillespie\cite{Gillespie_28,Gillespie_29,Melunis_30}. The algorithm determines the time step and reaction that would occur in an evenly distributed system accounting for random fluctuations. To implement the stochastic process, first, a cumulative reaction term $k_0$  is calculated as the sum of all the reactions $k_v$ occurring in the system, according to Eq (2). Using random numbers given by $r_1$ and $r_2$, the time step $\tau$ is added to the simulation time, and the reaction at that time step is chosen, according to Eq (3) and (4), respectively. 

\begin{equation}
k_0=\sum_{v=1}^{M}k_v
\end{equation}

\begin{equation}
\tau=(\frac{1}{k_0}ln\frac{1}{r_1})
\end{equation}

\begin{equation}
\sum_{v-1}^{\mu-1}k_v<r_2k_0\le\sum_{v=1}^{\mu}k_v
\end{equation}

To put it all together, given the initial conditions, the program calculates the reaction rate for the different processes: microwave absorption/emission, elastic/inelastic collision, and dissociation. Each molecule then starts with the initial internal energy. Based on the reaction chosen by Eq (4) and the time step by Eq (3), the appropriate energy is updated to the molecule’s internal energy, and the program progresses. This cycle repeats until the time exceeds the set limit or the molecule dissociates. The flowchart in Figure 1 demonstrates this process.

\begin{figure}
\centering
\resizebox*{7cm}{!}{\includegraphics{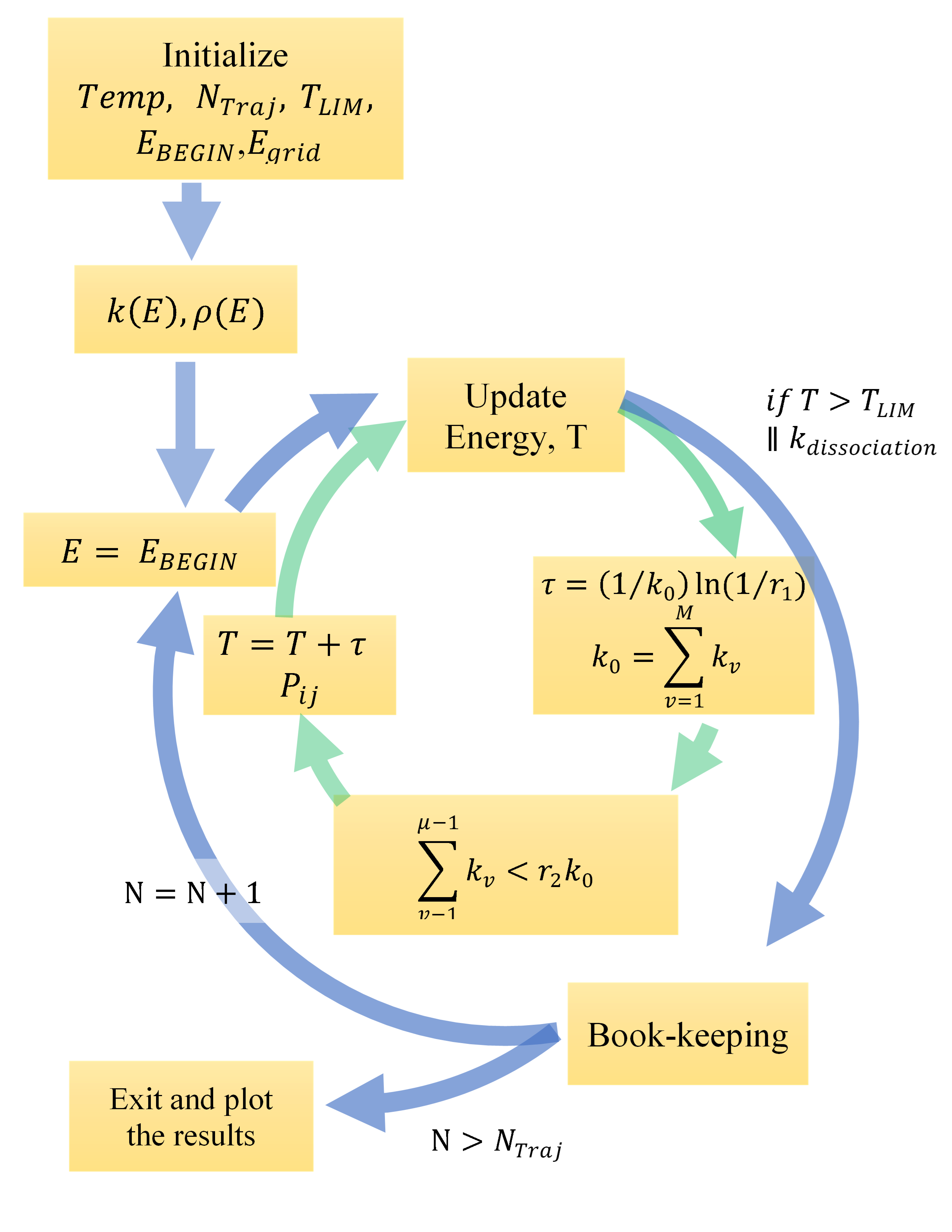}}
\caption{The flowchart for solving the master equation using Gillespie's algorithm. The system is initialized to the set parameters of temperature (Temp), Number of Trajectories (N$_{Traj}$), time limit (T$_{LIM}$), Initial energy (E$_{BEGIN}$) and size of energy grain(E$_{grid}$ reaction rates and density of states are initially calculated, and the stochastic process begins. Energy (E) is updated at each time step ($\tau$) and the simulation for each molecule runs if the set time has exceeded (T$>$T$_{LIM}$), or the molecule dissociates (k$_{dissociation}$).} 
 \label{Figure1}
\end{figure}

\subsection{EMF effects}
The electromagnetic wave interacts with a molecule by distorting the electron cloud in the direction of the applied field. As this distortion relaxes, energy is gained in the system due to intermolecular interactions leading to heating in the system. This mechanism is known as dielectric heating, and it primarily depends on a cluster of atoms or molecules, and it is not a quantum mechanical phenomenon.  

The property of the material that determines the coupling of electromagnetic waves is the relative permittivity. A material with permanent or induced dipoles will store charges if placed between two electrodes in a circuit, and the permittivity of the material defines this charge storing ability. The permittivity is defined as complex variable given in Eq(5).

\begin{equation}
\epsilon^*=\epsilon'-j\epsilon"
\end{equation}

Where $\epsilon'$ is the real part of the permittivity which is related to the charge storage and $\epsilon"$ is the imaginary part which indicates the loss term. As dielectric heating is known to heat the bulk of the sample, it is directly related to the loss tangent of the system. The loss tangent or the energy dissipation factor is given by Eq(6).

\begin{equation}
tan\delta=\frac{\epsilon'}{\epsilon"}
\end{equation}

The loss tangent can be experimentally calculated for different temperatures\cite{SEBASTIAN_31}. Dielectric heating can also be viewed as the efficiency of conversion of EMF energy into heat, and it is related to both the dielectric and thermal properties of the system. The relationship is given by Eq(7)\cite{Metexas_32}.  

\begin{equation}
P=\sigma|E|^2=(\omega\epsilon_0\epsilon'tan\delta)|E|^2
\end{equation}

Where P represents the power dissipated per unit volume in a material, $\sigma$ is the conductivity, E is the electric field and $\omega$ is the angular frequency. The power dissipated in the system is equivalent to the multiphoton jump due to the EMF absorption. In particular, we consider a bulk system with a continuum in the density of states. Therefore, modeling the EMF absorption as a multiphoton absorption is justified as the energy is assumed to be absorbed by all the molecules equally, i.e., we can relate this to the power absorbed by the whole system (not a single molecule). In the next step, the energy gained by the multiphoton absorption is dissipated quickly through collisional effects trying to maintain the equilibrium in the system. The thermal effect is seen as an increase in temperature, while the non-thermal effect is observed as a deviation from this equilibrium. 
 
The EMF intensity plays a crucial role in the dynamics of the system. At higher microwave intensity, voltage breakdown occurs leading to arc discharges and highly ionized gas plasma. Very high values of microwave power($>$10\textsuperscript{6}W/cm\textsuperscript{2}) can lead to a higher number of molecules in the excited states with insufficient time for redistribution of energy which can lead to wrong interpretation of the result.

\subsection{Molecular Collisions}
The collision frequency $\omega$ is a function of temperature, pressure, and atomic potential. The molecules are modelled as Lennard-Jones sphere and interact via the Lennard-Jones potential between molecules A and bath gas M. The collisional frequency is calculated using the following Eq(8)

\begin{equation}
\omega=\pi\sigma_{LJ}[M]\sqrt{\frac{8RT}{\pi\mu_{A-M}}}\Omega^{*}_{A-M}
\end{equation}

Where $\Omega^{*}_{A-M}$ is the collision integral given in Eq(9), [M] the number density of the system, $\sigma_{LJ}{M}$ the collision diameter, T(K) the temperature, and $\mu$ the atomic mass. The collision integral in this model is given by where ${\epsilon_{A-M}}$  is the depth of the Lennard Jones potential well\cite{Atkins_33,Troe_34}. 

\begin{equation}
\Omega \approx [0.636+0.567\log\frac{kT}{\epsilon_{A-M}}]
\end{equation}

A bi-exponential model gives the transfer probability function used in this model as in Eq (10).\cite{Troe_35,STREKALOV2010129_36}. 

\begin{equation}
P_{ij}=(1-a)e^{\frac{(E_i-E_j)}{\alpha}}+be^{\frac{(E_i-E_j)}{\beta}}
\end{equation}

Where a,b,$\alpha$, and $\beta$ are coefficients for collisional energy transfer. The up and down collision are related by the balance equation given by: 

\begin{equation}
\frac{P_{i,j}}{P_{j,i}}=\frac{\rho_i}{\rho_j}e^{(\frac{E_i-E_j}{k_bT})}
\end{equation}
The energy transfer, collision, and dissociation processes are all connected to the density of states. Many methods exist to calculate the density of states based on the size and accuracy of the molecule\cite{Beyer_37,Stein_38}. Statistical tools and computer programs are available to generate the density of states of a molecule based on the vibrational and rotational frequencies.

The system parameters, such as the energy distribution, photons absorbed /emitted, and trajectories reacted, are stored to evaluate the time evolution of the system. The time bin is fixed and stores the average value in that interval. This gives an advantage to the memory required for each simulation. Considering the system to be linear, the average values of an ensemble of molecules can describe the system's behavior. The size of the energy grain must be chosen to simulate the molecule with sufficient accuracy. Finer energy grain will result in more memory space and would be required in reactions with multiple channels.

\subsection{Unimolecular Dissociation}
A critical quantity to evaluate the reaction rate is the dissociation rate constant. The Rice-Ramsperger-Kassel-Marcus (RRKM) model is widely used to study the unimolecular kinetics and is based on the fundamental assumption that a micro-canonical ensemble of states exists initially for the activated molecule and is maintained as it decomposes. It is given by: 

\begin{equation}
k(E,J)=\frac{W(E,J)}{h\rho(E,J)}
\end{equation}
                
Where $W(E,J)$ is the transition state's sum of states. To include the EMF effect, we can assume that the system reaction to sudden changes is slow, and the average energy shifts as EMF energy is absorbed. Due to the non-equilibrium distribution of the system, the vibrational temperature of the system would be different. This can be considered a change in its vibrational temperature and be included empirically in the RRKM equation, according to Eq (13).

\begin{equation}
k_{MW} = Ae^{\frac{-E_{crit}}{\beta(E)T}}
\end{equation}

Here A and $\beta$ values are determined experimentally by fitting the plot of the observed values.

\section{Results}

EMReact is written in MATLAB, and the source code is available and can be downloaded from the provided link\cite{Code_39}.  The code, without the EMF inclusion, is written based on the MULTIWELL suite developed by Barker et. al.\cite{Barker_40,Barker_41,BARKER_42,Code_43}. The program utilizes MATLAB functionalities and offers graphical visualization and prompt inputs for a user-oriented approach. 

\begin{figure}
\centering
\resizebox*{6cm}{!}{\includegraphics{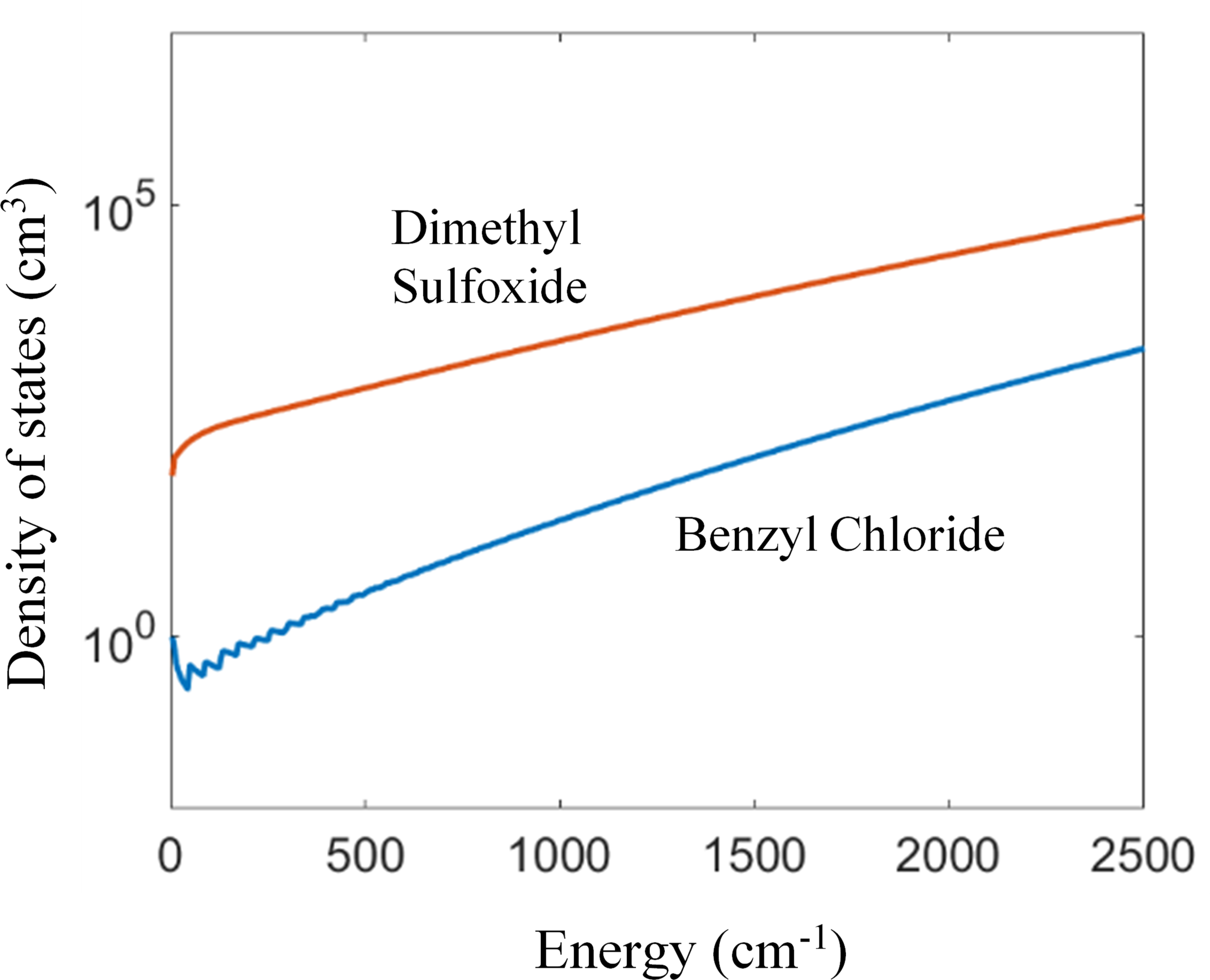}}\hspace{5pt}
\caption{Density of states for DMSO and Benzyl Chloride} 
 \label{Figure2}
\end{figure}
For this work, we have chosen two systems which that are known to exhibit thermal as well as non-thermal effects. Dimethyl Sulfoxide which is a solvent and good EMF absorber, and Benzyl Chloride which is a precursor to several microwaves-assisted reactions. The density of states is calculated for an energy grid of 5 cm$^{-1}$ employing the Wang-Landau method implemented in the Desnum package of the MULTIWELL suite, as shown in Figure 2. The density of states, as shown, is relatively smooth even at lower energy which leads, leading to little error during interpolation for the selected energy grid. The vibrational frequencies used to calculate the density of states for the two systems are calculated and experimentally determined by Chen and Naganathappa\cite{Chen_56, Naga_56}.

\subsection{Dimethyl Sulfoxide}
Dimethyl Sulfoxide (DMSO) is an aprotic polar solvent that has been used extensively in microwave-assisted processes\cite{TIAN2020110523_44,C5RA18039B_45}. DMSO can couple to the microwave field increasing its temperature due to dielectric losses which makes it one of the reasons it is widely used. The parameters used for the simulations are given in Table 1.

\begin{table}
\tbl{Simulation Parameters for DMSO\cite{DMSO_46}}
{\begin{tabular}{c} \toprule
 $\sigma_{LJ}$=3.66, $\epsilon_{A-M}=168.18$\\
 $tan_\delta$=0.8524, $\epsilon$=13, $\omega$=2.4 GHz\\
 N$_{TRAJ}=3000$, E$_{BEGIN}$=1000 cm$^{-1}$,  T$_{TLIM}$=5 ms, Temp=300 K
\end{tabular}}
\label{Table 1}
\end{table}

\begin{figure}[hb!]
\centering

\includegraphics[scale=0.9]{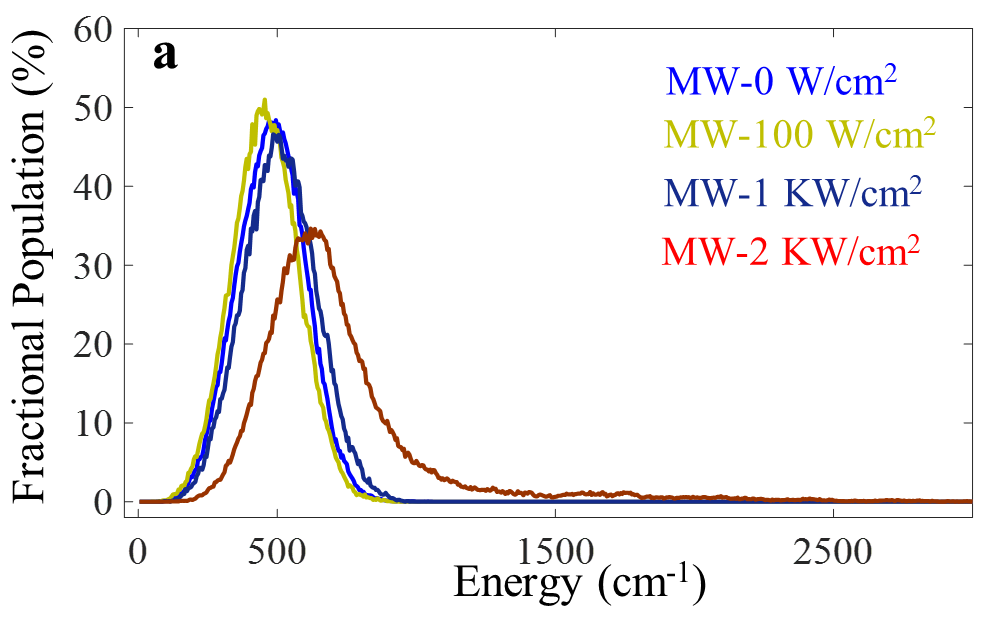}%
\includegraphics[scale=0.34]{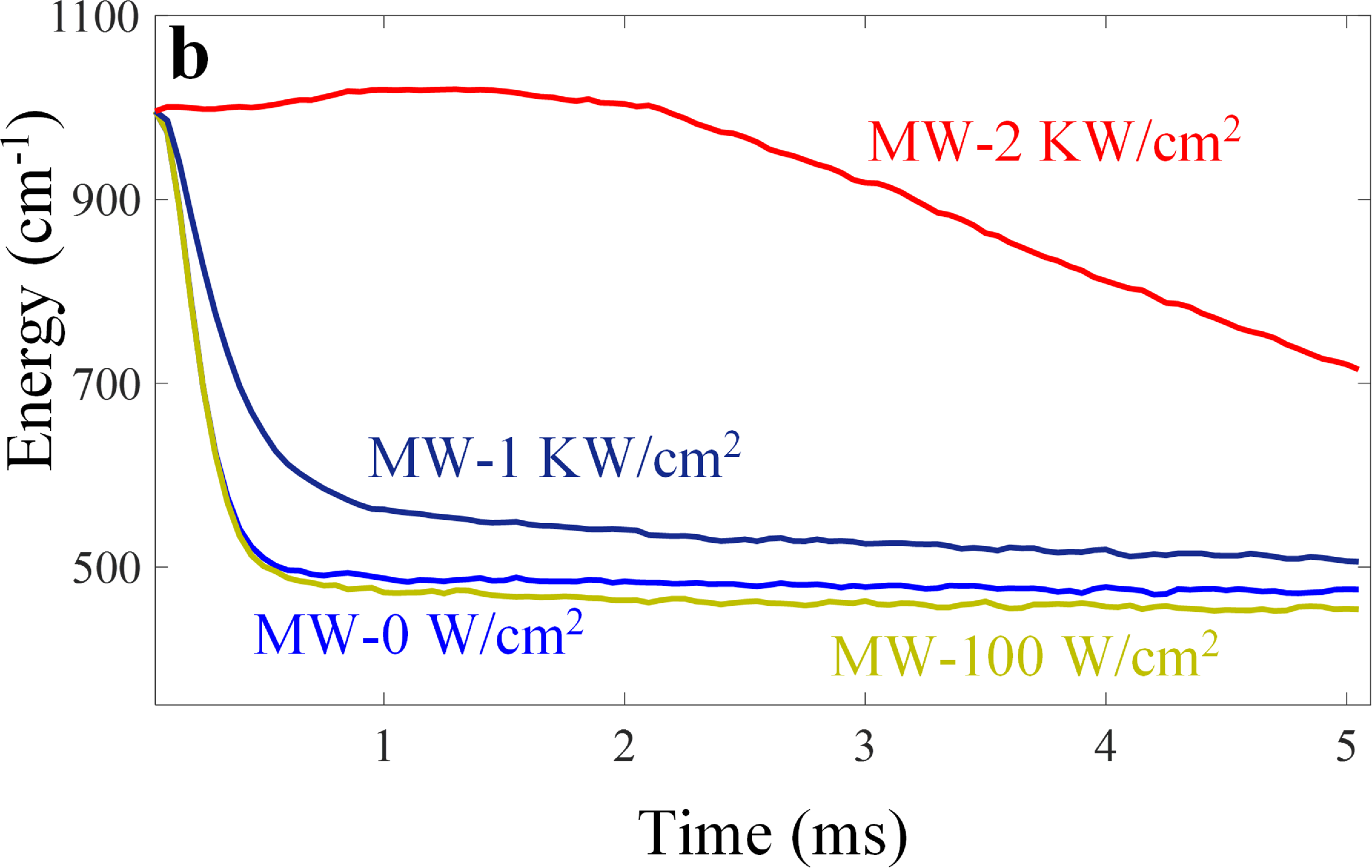}%
\caption{a: Internal Energy distribution for DMSO at 5ms. b: Average energy $<E>$ with respect to time} 
 \label{Figure3}
\end{figure}

The internal energy distribution at different microwave intensities for a time limit of 5 ms is given in Figure 3(\textit{a}). Without microwave radiation (MW=0W/cm$^2$) the internal energy follows a Boltzmann distribution given by $Ae^{(-(E-Ei)/k_B T)}$. The starting internal energy set at 1000 cm$^{-1}$ reaches equilibrium due to the vibrational-vibrational energy transfer by the collisional process in 500 $\mu$s. This is shown by the average energy plot shown in Figure 3(\textit{b}). The average energy $<E>$ of the internal energy distribution depends on the temperature of the system.  As microwave radiation is added to the system, the average energy plot shifts based on the microwave intensity. 

At the microwave power of 100 W/cm$^2$, the energy distribution is modified such that the trailing edge of the distribution has a lower slope than the leading edge indicating a non-uniform distribution. Unequal distribution is difficult to observe on the internal distribution plot but can be seen by the average energy plot. The thermal effect in the form of shifting of the distribution due to temperature increase is not observed due to the time limit of the simulation. The average energy plot for 100 W/cm$^2$ is less than the average energy at no microwave radiation due to the spread of the energy state in the trailing edge of the energy distribution.

\begin{figure}
\centering
\resizebox*{6cm}{!}{\includegraphics{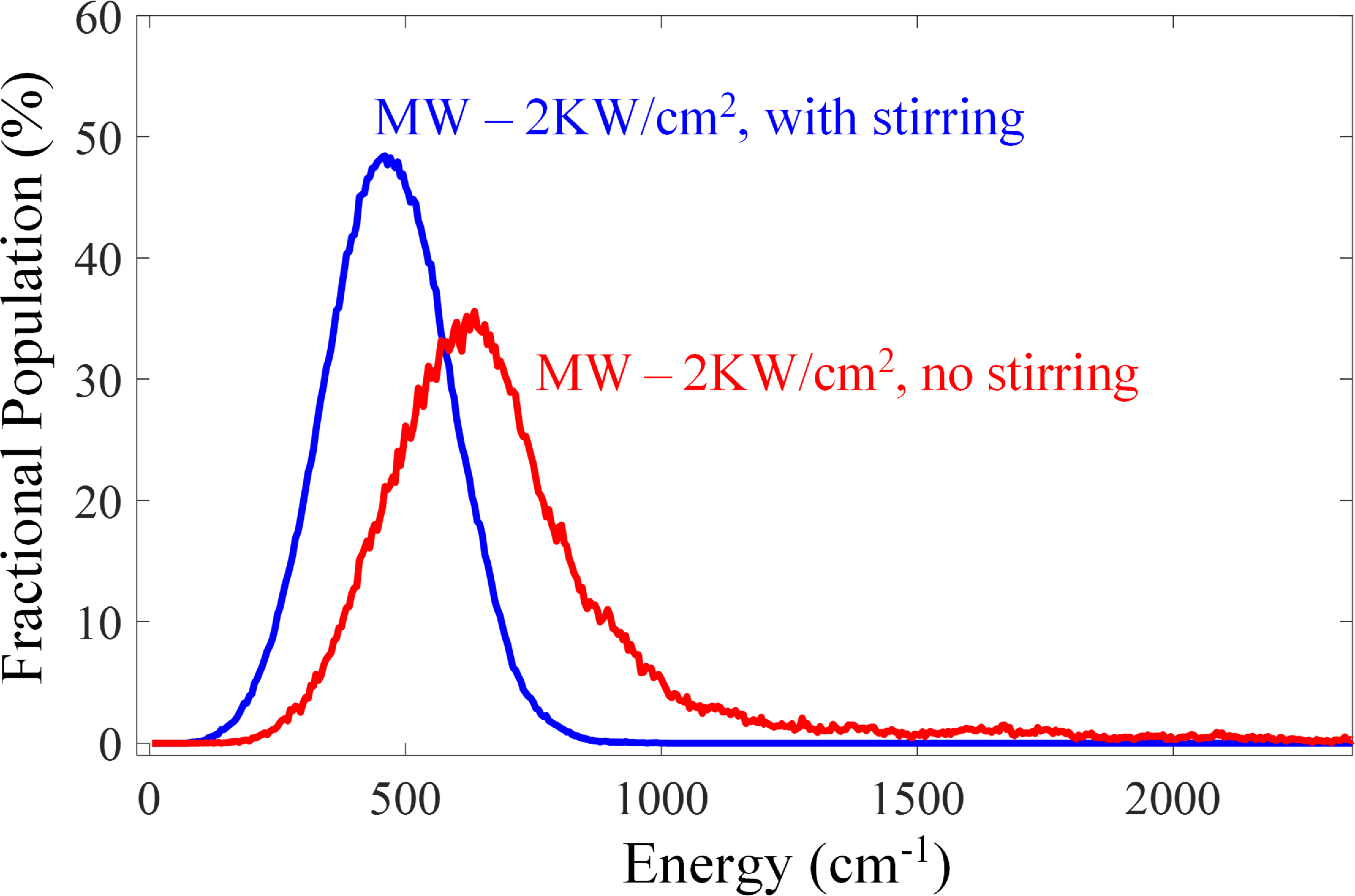}}
\caption{Internal energy distribution by microwave radiation at 2 kW/cm$^{2}$ with and without stirring. The collisional effects distribute the energy more efficiently thus reducing the sudden increase in temperature.} 
 \label{Figure4}
\end{figure}

At higher microwave power at 1 kW/cm$^2$, the thermal and non-thermal effect is evident through the shift in the energy distribution and the slight spread in the population at high energy states. The higher energy states are more susceptible to microwave absorption due to the continuum of the density of states compared to the lower states. The average energy plot is also shown to be shifted higher than the equilibrium distribution. 

An important factor in many microwave experiments is the effect of stirring the system under radiation. It has been observed that the microwave effect is reduced under stirring\cite{Herrero2007-vx_12}. The effect of stirring can be incorporated into the system by increasing the collisional frequency. The vibrational-vibrational energy transfer is responsible for redistributing the energy in the system leading to equilibrium at a fixed temperature. Increasing the collisional frequency increases this redistribution and effectively cancels the microwave effect. As shown in Figure 4, the internal energy distribution for a system under 2 kW/cm$^2$ of microwave power represents a thermal equilibrium distribution after stirring.

\section{Benzyl Chloride}
Microwave-assisted synthesis of phosphonium salt through the nucleophilic substitution of benzyl chloride with triphenylphosphine has been reported and has shown to exhibit non-thermal effects at 373 K\cite{Cvengros_47}. Benzyl chloride is a relatively neutral to microwave absorption with a low loss tangent. The simulation parameters are shown in Table 2.

The internal energy distribution at a time limit of 5 ms is shown in Figure 5(\textit{a}) at different MW intensities. The corresponding average energy plot with respect to time is shown in Figure 5(\textit{b}). At low microwave intensity, the shift in the equilibrium is not observed. As Benzyl Chloride is not a good absorber of MW, the temperature changes through the shift in the equilibrium distribution under microwave radiation are not prominent for the time limit of 5 ms.  The internal distribution at a microwave power of 100 W/cm$^2$ and 1 kW/cm$^2$ does not show many changes compared to the distribution with no Microwave Power. The average energy plot shows a shift in the energy level compared to the reference energy indicating a non-thermal effect.

\begin{table}
\tbl{Simulation Parameters for Benzyl Chloride\cite{BenzC_48}}
{\begin{tabular}{c} \toprule
 $\sigma_{LJ}$=6, $\epsilon_{A-M}=410$\\
 $tan_\delta$=0.08, $\epsilon$=7, $\omega$=2.4 GHz\\
 N$_{TRAJ}=3000$, E$_{BEGIN}$=1000 cm$^{-1}$,  T$_{TLIM}$=5 ms, Temp=350 K
\end{tabular}}
\label{Table 2}
\end{table}

The non-thermal effect, seen as a nonequilibrium distribution, is caused by some states pushing to higher energies while the collisional reaction constantly tries to redistribute this change, resulting in a longer energy trail in the higher energy side with a shift in the peak of the distribution.

\begin{figure}
\centering
\includegraphics[scale=0.34]{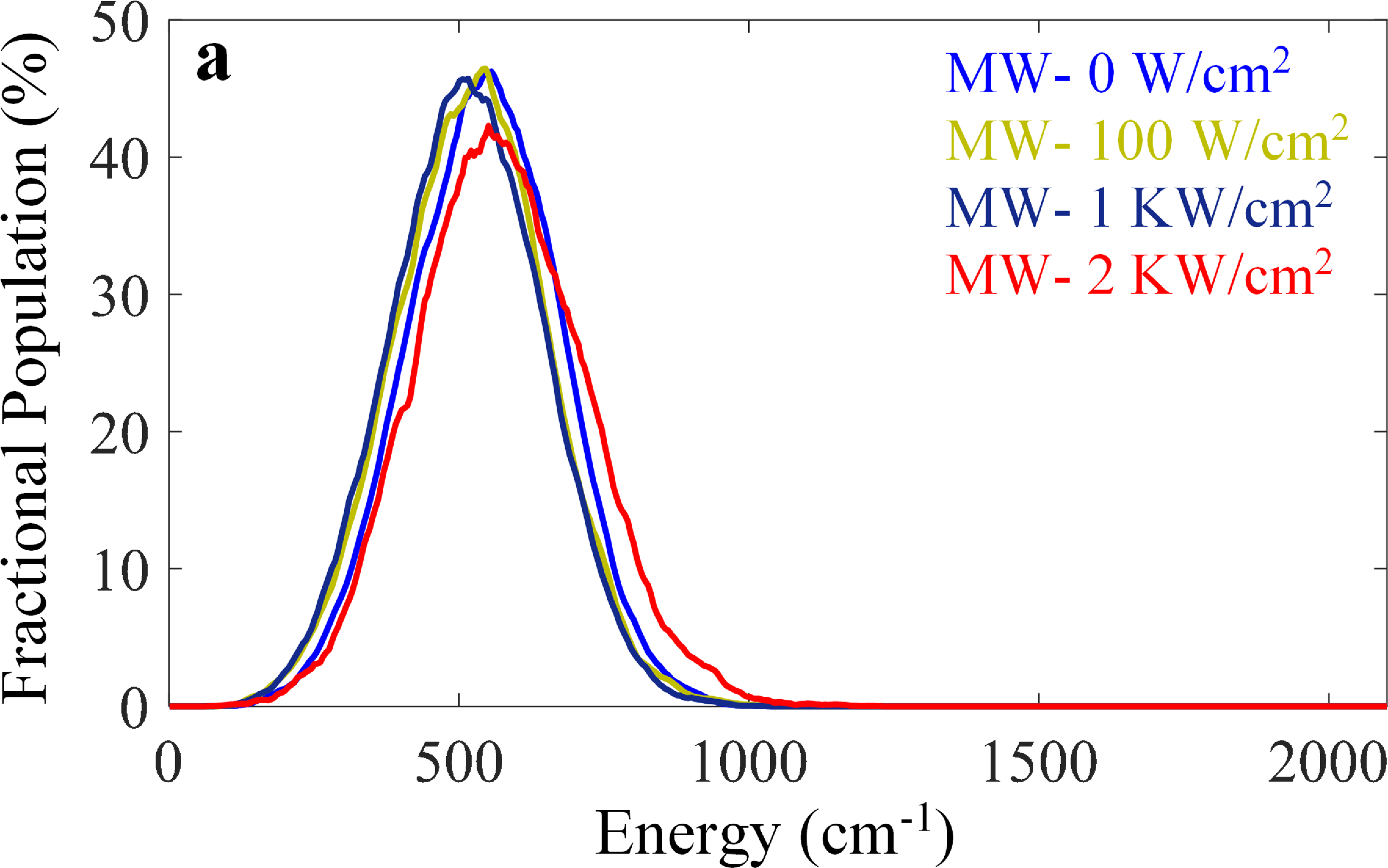}%
\includegraphics[scale=0.34]{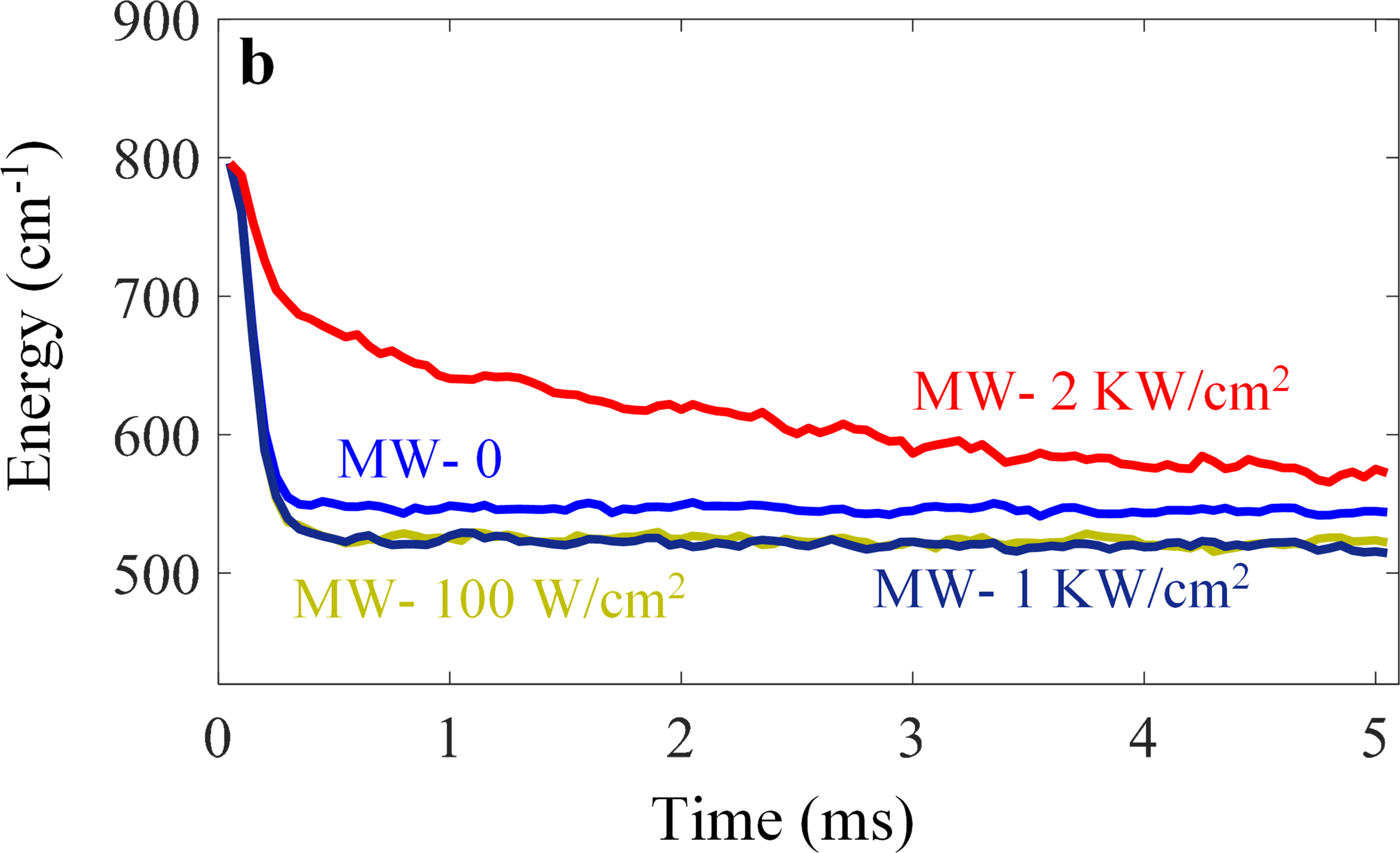}%
\caption{a: Internal energy distribution for Benzyl Chloride at 5ms. b: Average energy $<E>$ with respect to time} 
 \label{Figure5}
\end{figure}

Based on the simulation parameters and the competing microwave absorption/emission, there is a microwave intensity beyond which the non-thermal effect will be more evident, which we call the characteristic intensity. The characteristic intensity can be defined as the intensity beyond which the microwave absorption/emission rate dominates over the collision rate. It is expected that microwave induces a nonequilibrium distribution at this level and exhibits a non-thermal effect. Figure 6 shows the microwave absorption/emission rate of the two reactants studied, along with the collisional frequency. The rate of microwave absorption/emission increases linearly with microwave power. It is determined by the permittivity and the density of states of the system; hence, it is a material-dependent value. The microwave characteristic intensity for DMSO is found to be 2 kW/cm$^2$, and for Benzyl Chloride is 15 kW/cm$^2$. 

\begin{figure}[h]
\centering
\resizebox*{7.5cm}{!}{\includegraphics{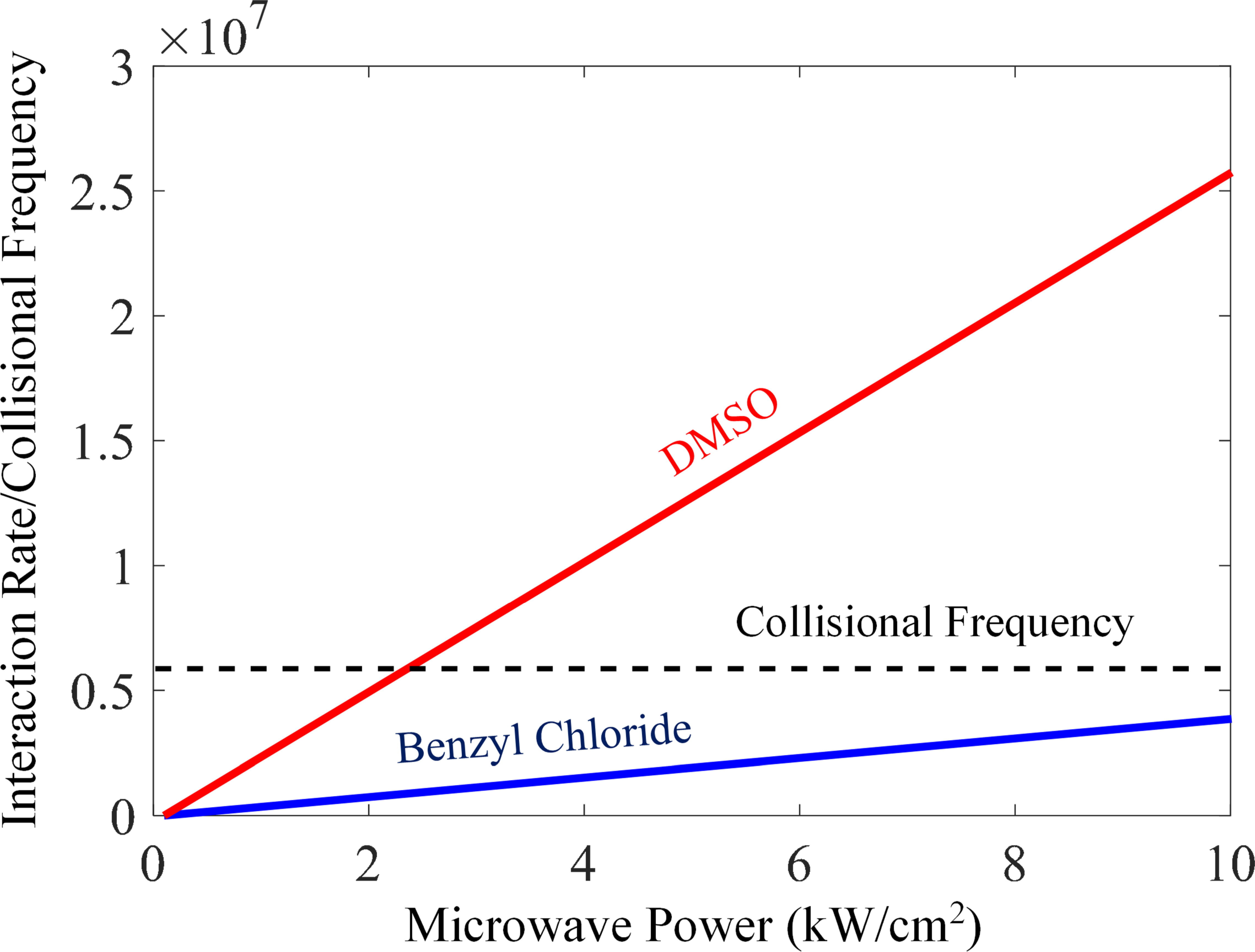}}
\caption{Interaction rate and collisional frequency versus microwave intensity for reactants (DMSO: red, Benzyl chloride: blue) along with the average collisional frequency(black).} 
 \label{Figure6}
\end{figure}

\subsection{Dissociation Reactions}
In studying the enhancement of reaction rate under the microwave, we have implemented unimolecular dissociation simulations with CO$_2$ and H$_2$O$_2$ molecules in an argon bath under microwave radiation. The microwave effect on CO$_{2}$ and H$_2$O$_2$ has been studied and reported for applications in lasers, carbon monoxide conversion, microwave plasma/catalyst and waste-water treatment\cite{Liao_49,LIAO_50,Hunt_51,Bekerom_52,Qin2018_53}.  
The dissociation rate constant for CO$_2$ is given in the literature\cite{Ibragimova2000_54,etde_6336464_55}. Figure 7(\textit{a}) presents the plot of trajectories reacted with increasing microwave intensity. Following the trend in the figure, the reaction rate increases with both temperature and microwave intensity. At low MW intensities, the reaction rate is too low for temperatures below 1200 K. However, when the MW intensity increases above 10$^4$ W/cm\textsuperscript{2}, the reaction rate becomes significant even at 1000 K. At high temperatures, the reaction rate saturates at lower MW intensity. For example, at 1500 K, the rate slows down at about 5×10$^4$ W/cm\textsuperscript{2}, while at 1000 K, the rate change happens at about 2×10$^5$ W/cm\textsuperscript{2}. The main reason for this difference is that at higher temperatures, the collision rate dominates and eventually spreads the distribution over the MW absorption and emission; hence, the effect of MW becomes less significant. 

\begin{figure}[h]
\centering

\includegraphics[scale=0.9]{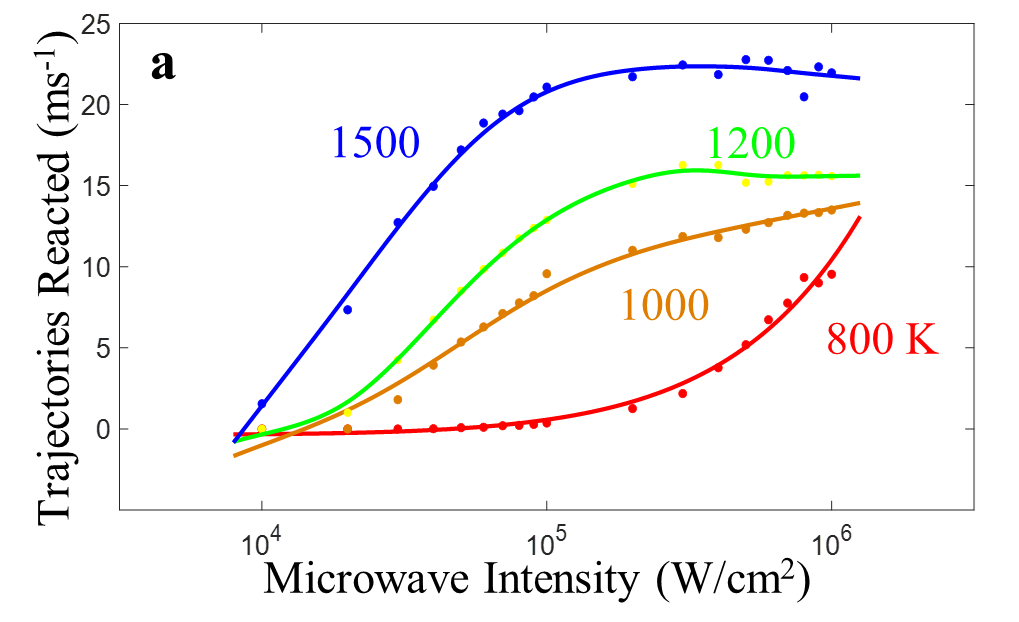}%
\includegraphics[scale=0.9]{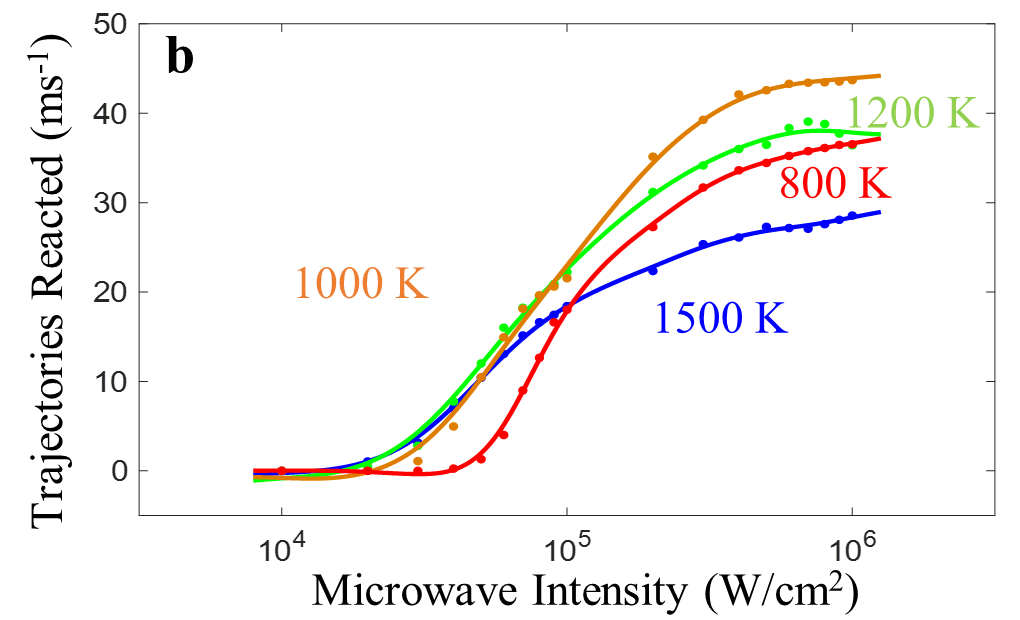}%
\caption{ a: Reaction rate for CO$_2$ versus microwave intensities at different temperatures. b: Reaction Rates for H$_2$O$_2$ for increasing microwave intensities at different temperatures.} 
 \label{Figure7}
\end{figure}

For unimolecular dissociation of H$_2$O$_2$, the trajectories reacted with respect to microwave intensity are shown in Figure 7(\textit{b}). The slope of the reaction rate increases initially with increasing temperature and there is a downward trend at a temperature higher than 1000 K. The difference in the plot for these two molecular systems can be attributed to the difference in the density of states and dissociation rate constant. At higher temperatures, the collisional effects in the H$_2$O$_2$ molecule play an essential role in the redistribution of energy attained from multi-photon absorption. 
Reducing the reaction rate at a higher temperature may seem counter-intuitive, but the similar argument of energy dissipation due to collision is the reason for this effect. The collisional parameters at higher temperatures dominate and due to the increase in the accessible density of states at high temperatures leads to faster energy dissipation from photon absorption and a reduction in the reaction rate. 

\section{Conclusion}

The discussed model gives a realistic description of the temporal evolution of the population due to multiphoton absorption in the presence of collision. Solving the master equation using a Monte Carlo approach can be used to solve several problems related to EMF-assisted reactions. Through the examples of DMSO and Benzyl Chloride, an idea of how we can differentiate between the thermal and non-thermal effects under EMF radiation is discussed.

Educational insight into stirring leading to the reduction of the EMF effect is explored as a collisional effect. The thermal effect under EMF radiation can be seen as a shift in the internal energy distribution due to the dielectric loss leading to an increase in the temperature of the system. The non-thermal effect was shown as a deviation from the equilibrium distribution with an unequal number of molecules in the high and low energy states.


\section*{Acknowledgement(s)}

This study is partially based upon work supported by the Air Force Office of Scientific Research (AFOSR) under contract number FA9550-19-1-0363 and the National Science Foundation (NSF) under grant number CBET-2110603

\section*{Disclosure statement}

The authors report there are no competing interests to declare.

\section *{Data availability statement}

The data that support the findings of this study are openly available at https://github.com/keludso/EMReact.git  and https://research.ece.ncsu.edu/nanoengineering \cite{Code_39}.


\end{document}